\documentclass{elsart3}
\usepackage{amssymb}
\usepackage{graphicx}

\def\aprle{\buildrel < \over {_{\sim}}} 
\def\aprge{\buildrel > \over {_{\sim}}} 
\def\nubar{\overline{\nu} }
\def\ldens{\mathcal{L}}
\def\lum0{\left ( L_\odot ~{\rm Mpc}^{-3} \right) } 
\begin{document}

\begin{frontmatter}

\title{Perspectives of High Energy Neutrino Astronomy}

\author{Paolo Lipari}

\address{INFN and Dipartimento di Fisica, Universit\`a di Roma 1}

\begin{abstract}
This work  discusses the  perspectives
to  observe  fluxes of
high energy astrophysical neutrinos  with
the planned  km$^3$  telescopes.
On the basis of  the observations of  GeV and TeV  $\gamma$--rays, and
of  ultra  high energy cosmic  rays, 
it is possible to construct     well motivated 
predictions   that indicate  that
the discovery of such  fluxes is  probable.
However the range of  these predictions  is   broad, and  the 
very important opening of the ``neutrino  window'' on the 
high energy universe is not  guaranteed with the current design
of the detectors.
The problem of  enlarging the 
detector  acceptance using   the  same  (water/ice Cherenkov) 
or  alternative (acoustic/radio)  techniques is  therefore of  central 
importance.
\end{abstract}

\end{frontmatter}

\section{Introduction}
\label{sec:introduction}
The fundamental  significance of the opening
of the new ``window''   of 
high energy $\nu$  astronomy\footnote{Neutrino Astronomy for solar
and gravitational collapse SuperNovae  $\nu$  has already been
remarkably successful. This discussion  will only consider
high energy ($E_\nu \aprge 1$~GeV)  neutrinos.}
 for the 
observation of the universe is beyond discussion.
Neutrinos  have properties that are 
profoundly  different  from those of  photons, 
and observations  with this  new  ``messenger''  will  allow
us to  develop a  deeper  understanding    of  
known astrophysical  objects, and
 also likely lead to the
discovery of  new unexpected  classes of
sources, in the same way as the
opening of each new window  in  photon  wavelength has  lead
to  remarkably  interesting  discoveries. 

If there are few  doubts that in the future
neutrino astronomy   will mature  into  an essential
field  of observational 
astrophysics, it is less clear how long and difficult
the history of  this  development will be,
and in particular  what will be the 
scientific  significance  of the results of the planned
km$^3$  telescopes.
These telescopes are
``discovery'' instruments and 
only {\it a posteriori},  when their data  analysis 
is completed,  we will be able to appreciate
their  scientific  importance.
It is   however interesting to  attempt
an estimate, based on our present  knowledge,
of the expected  event rates   and number  of  detectable  sources.

As a warning,  it can be amusing   to recall 
that when (in june 1962)  Bruno Rossi, Riccardo Giacconi and  their
colleagues  flew the first $X$--ray telescope  \cite{giacconi,giacconi-first} 
opening the X--ray  photon  window  to observations,
the  most promising  $X$--ray  source  was the sun,
followed  by the moon
(that could  scatter the solar wind).
It is now known 
that the moon  surface does indeed emit  X--rays,
but in fact it appears  as a dark  shadow  in the $X$--ray  sky, 
because it eclipses  the emission  of 
the  ensemble of  Active Galactic  Nuclei (AGN)  that 
can now be detected with a density of $\sim 7000$ per square degree
\cite{Brandt-Hasinger}.
So the $X$--ray  sky was  very generous to the observers,
and embarassed the theorists  who had worked  on
{\it a priori}  predictions.

The lesson here  is  that predictions   about the unknown are difficult.
One should expect   the  unexpected, and 
it is  often wise to take (calculated)  scientific  risks.
We all hope  that  history will repeat itself,
and that also the $\nu$  sky  will 
be generous to the brave scientists who, with 
great effort,  are constructing
the  new  telescopes.

On the other hand, we are  in the position 
to make  some  rather   well
motivated   predictions  about
the  intensity of   the astrophysical  $\nu$ sources,
because  high energy  $\nu$  production is 
intimately related to    $\gamma$--ray  emission
and cosmic  ray (c.r.)  production, and
the observations of  cosmic rays and  high energy
(GeV and TeV  range)   photons 
do give  us   very important  guidance
for the prediction of the neutrino  fluxes.

This review will  concentrate only on  high energy
astrophysics.  There are other  important scientific
topics  about  km$^3$ telescopes  that will not be covered here.
A  subject of comparable importance is ``indirect'' search for 
cosmological Dark Matter via the observations of
high energy $\nu$   produced at the center of the sun and the
earth  by the annihilation of DM  particles \cite{Bertone}.
The km$^3$ telescopes  can also be used to look for 
different ``New Physics'' effects such as 
the existence of Large Extra dimensions \cite{Halzen:2006mq}.
The  instruments also have the potential to perform
interesting  interdisciplinary studies.

\section{Components of the Neutrino Flux}
\label{sec:sources}
The observable neutrino  flux
can be schematically written as the  sum
of several components:
\begin{eqnarray}
\phi_{\nu_\alpha} (E, \Omega) & = &
\phi_{\rm atm}^{\rm standard} (E, \Omega)  +
\phi_{\rm atm}^{\rm prompt} (E, \Omega)  
\nonumber \\[0.2cm]
& + &  \phi_{\rm Galactic} (E, \Omega)  +
 \phi_{\rm Extra~Gal} (E, \Omega)   
\nonumber \\[0.2cm]
 & + &
\sum_{\rm Galactic} \phi_j  (E) ~ \delta [\Omega - \Omega_j] 
\nonumber  \\[0.2cm]
& + & \sum_{\rm Extra~Gal} \phi_k  (E) ~ \delta [\Omega - \Omega_k] 
\nonumber
\end{eqnarray}
The first two components  describe 
atmospheric  neutrinos 
that are generated 
 in  cosmic ray showers in the Earth's  atmosphere.
They are  an important foreground  to  the more
interesting  observations of the astrophysical components.
Atmospheric  $\nu$'s
can be split into  two components,   the first ``standard''
one is   due to the decay of charged  pions and kaons,
while the second one is due to the weak decays 
of short lived  (hence ``prompt'') 
particles  containing heavy quarks,
with charmed particles  accounting for
essentially all of the flux.
The prompt  contribution is   expected to be dominant
in the atmospheric  neutrino fluxes 
at high  energy. This  component has  not yet
been identified, and its prediction is  
significantly more  difficult   than  the standard flux,
because of our poor knowledge of the dynamics 
of charmed  particles production
in hadronic interactions.
\begin{figure}[ht]
 \includegraphics[width=5.2cm,angle=90.]{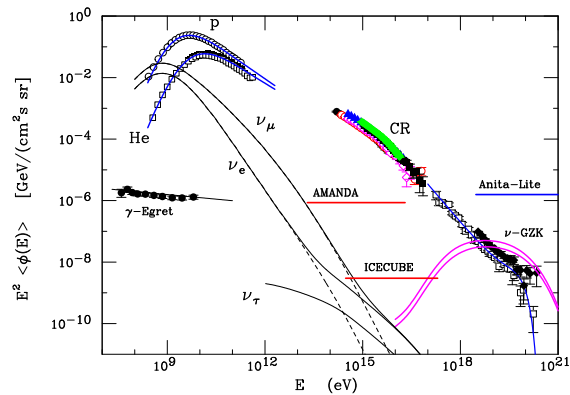}
   \caption{\footnotesize 
Neutrino fluxes.  The solid lines  labeled $\nu_e$, $\nu_\mu$ and
$\nu_\tau$  show  the  angle averaged
 fluxes of  atmospheric  neutrinos and anti--neutrinos  
as a function of energy. The associated 
dashed lines are   the fluxes
neglecting the  ``prompt''  contribution of  charmed 
hadrons. The  lines labeled $\nu$--GZK  describes
GZK neutrinos  \cite{ESS}.
The points  (and  fitted  lines)  describe CR  direct and
indirect measurements.  The horizontal lines  labeled
Amanda and  Icecube are the current limit and the predicted 
sensitivity of km$^3$  $\nu$ telescopes for an isotropic
$\nu$ flux. 
The line  labeled  Anita--Lite is the limit on the $\nu$ flux
obtained with radio methods \cite{anita-lite}.
The  isotropic   $\gamma$--ray  flux  measured by  Egret
\cite{egret-extra} is also shown.
\label{fig:cr_nu}  }
\end{figure}
Figure~\ref{fig:cr_nu} shows  the angle  averaged
atmospheric  $\nu$ fluxes \cite{bartol} 
with an  estimate of the ``prompt''  component \cite{lip} 
that  ``overtakes'' the standard one at 
$E_\nu \sim 10$~TeV for $\nu_e$ and $\sim 100$~TeV for
$\nu_\mu$.

The next  two components 
of the neutrino fluxes 
are ``diffuse''  fluxes  coming from  $\nu$   production
in interstellar space in our own Galaxy, and in
intergalactic  space.
The  diffuse Galactic  emission is  due to the interaction
of cosmic rays  (confined inside the Milky Way by  the
 galactic  magnetic fields)
with the gas present in interstellar  space.
The  angular  distribution of this emission is  expected
to be concentrated in the galactic  plane,   very likely with a
distribution similar to the one observed for
GeV photons by EGRET \cite{egret-gal}.
The extragalactic  diffuse  emission is 
 dominated by the decay
of  pions  created in  $p\gamma$ interactions  by
Ultra High Energy protons ($E_p \aprge 6 \times 10^{19}$~eV)
interacting with the 2.7K cosmic  radiation.
These ``GZK''  neutrinos are  present only at very  high energy
(with the flux peaking at $E_\nu \sim 10^{18}$~eV).

Together with the diffuse fluxes  one  expects the   contribution
of an ensemble of  point--like (or quasi--point like) 
sources  of galactic  and  extragalactic  origin.
Neutrinos  travel along straight lines  and allow the imaging
of these sources.
It is  expected that most of the  extragalactic  sources 
will  not be resolved, and therefore the  
ensemble of  the 
extragalactic point sources
(with the exception of  the closest and brightest sources)
will appear as a diffuse, isotropic flux
that can 
in principle be  separated from 
the atmospheric   $\nu$ foreground because
of a  different   energy spectrum, and flavor composition.

\section{Neutrinos, Photons  and Cosmic Rays} 
\label{sec:nuhadgam}
In the standard mechanism  for  the production of
high  energy $\nu$  in astrophysical  sources
a population of  relativistic  hadrons  (that is cosmic  rays)
interacts with a target (gas or radiation fields) 
creating  weakly decaying  particles 
(mostly $\pi^\pm$, and Kaons)  that produce $\nu$
in their decay.
The  energy spectrum  of the  produced  neutrinos
obviously  reflects the spectrum  of the parent cosmic rays.
A  well known  consequence of the  approximate 
Feynman   scaling of the   hadronic interaction  inclusive
cross sections  is the fact   that if the parent c.r. 
have a power  law spectrum of form
 $\phi_{\rm cr} \simeq K_{\rm cr} \;E^{-\alpha}$,
and their  interaction probability is  energy independent\footnote{
In the most general case c.r. of different
rigidity $p/Z$  diffuse in  different ways  inside the source
and  have  different space  distributions.
For a non homogeneous   target, this can be reflected in an  energy 
dependent interaction  probability.}
the  $\nu$ spectrum, to a good approximation,
is   also  a power law  of the same  slope. 

The  current favored models for  1st order Fermi acceleration 
of  charged  hadrons near astrophysical  shocks 
predict  a   generated spectrum with a slope  $\alpha \simeq 2$.
This  expectation is in fact confirmed  by the observations
of young SNR by  HESS, and leads the expectations
that   astrophysical $\nu$  sources
are also likely to  have power law spectra with slope close to 2.
Such a slope 
is also predicted in  Gamma  Rays Bursts models \cite{Waxman-GRB}
where   relativistic  hadrons interact  with a
power law  photon  field.
It is  important 
to  note that the high energy cutoff  
of the parent c.r. distribution  is   reflected in
a  much more gradual  steepening of the  $\nu$  spectrum 
for  $E_\nu  \aprge 0.01~E_{p, {\rm max}}$.

\begin{figure}[ht]
 \includegraphics[width=5.15cm,angle=90.]{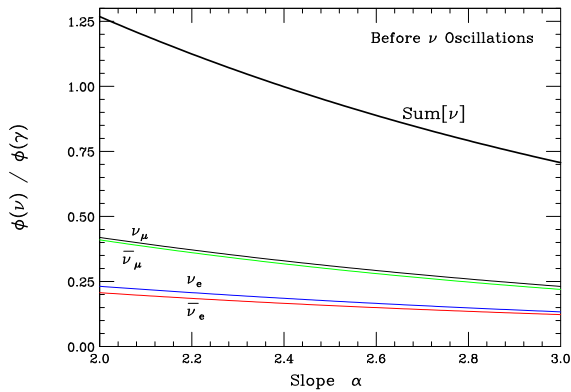}
   \caption{\footnotesize 
Ratio $\phi_\nu/\phi_\gamma$  between the  neutrino  and 
photon fluxes at $E \sim 3$~TeV, 
for a power law c.r. population  of slope $\alpha$
and p/n ratio of 10\% interacting with a low density
gas  target.
Photon absorption is neglected. 
The calculation uses
the Sibyll \cite{sibyll} hadronic interaction model.
The thick line sums over all neutrino type.
The  $\nu$ flavor is 
calculated at the source  before the inclusion of oscillations.
\label{fig:ratio}  }
\end{figure}

The hadronic interactions  that are the sources  of 
astrophysical neutrinos also create  a  large  number of
$\pi^\circ$  and $\eta$  particles that 
decay  in a $\gamma\gamma$ mode   generating high energy
$\gamma$--rays.  In general, it is possible 
that these photons are  absorbed 
inside the source,  and the energy  associate  with them can
emerge at lower   frequency, however for  a transparent source
the relation  between the  photon and  neutrino fluxes
is   remarkably robust. 
For  a power law c.r.  spectrum,
the $\gamma$--rays  are  also 
created  with  a  power law  spectrum of  the same slope.
The  approximately constant $\nu/\gamma$ ratio
is  shown  in fig.~\ref{fig:ratio}
as a function  of the slope $\alpha$.
The  $\nu/\gamma$  ratio
is approximately constant in energy
when $E$ is much smaller than the c.r. energy cutoff, and 
depends only weakly  on the hadronic  interaction model.

At the source, 
the relative  importance
of the  different   $\nu$  types is:
$$
\{\nu_e,\overline{\nu}_e, 
\nu_\mu,\overline{\nu}_\mu, 
\nu_\tau,\overline{\nu}_\tau \} ~ \simeq  ~
\{1+\epsilon,1-\epsilon,2,2,0,0\}
$$
where $\epsilon \sim 0.1$  depends
on the slope  of the  spectrum,  the relative importance
of protons  and heavy nuclei in the c.r.  parent population,
and the nature of the c.r. target (gas or radiation field).
The ratio $\nu_\mu/\nu_e \simeq 2$  at the source 
is a well known  consequence of the fact that the chain  decay of a
charged pion: $\pi^+ \to \mu^+ + \nu_\mu$
followed by
$\mu^+ \to e^+  + \nu_e + \nubar_\mu$
generates two $\nu$   of $\mu$--flavor and one of
$e$--type. The presence of a $\nu_\mu$ and an 
$\overline{\nu}_\mu$ in the final state  insures
that    the ratio  $\nu_\mu/\overline{\nu}_\mu \simeq 1$,
while the ratio  
 $\nu_e/\overline{\nu}_e$ is controled  by the
relative  importance  of $\pi^+$  over $\pi^-$ production
that is not symmetric for a $p$ rich c.r. population.

\section{Neutrino  Oscillations}
Measurements of solar and atmospheric neutrinos  have  recently
established the existence of neutrino oscillations,
a quantum--mechanical   phenomenon  that is 
a consequence
of the non--identity of the  $\nu$  flavor
\{$\nu_e$, $\nu_\mu$, $\nu_\tau$\} and
mass \{$\nu_1$, $\nu_2$, $\nu_3$\} eigenstates.
A  $\nu$    created with  energy $E$ and  flavor $\alpha$  
can  be  detected   after a distance $L$ with a different
flavor $\beta$ with  a probability that  depends 
periodically  on the ratio $L/E$.
This  probability oscillates  according  to three  frequencies
that  are proportional to the  difference between 
the  $\nu$  squared masses
$\Delta m^2_{jk}$, and different  amplitudes  related
to the mixing  matrix $U_{\alpha j}$ that relates the flavor and
mass  eigenstates.
The  shortest (longest) oscillation   length,
corresponding to the 
largest  (smallest) $|\Delta m^2|$)  can  be written as:
\begin{eqnarray*}
\lambda_{12} & = &  (4\pi \,E_\nu) /| \Delta m^2_{12} |
 \simeq 3.1 \times 10^{12} ~E_{\rm TeV} ~ ~ {\rm cm},
\\
\lambda_{23} & = &  (4\pi \,E_\nu) /|\Delta m^2_{23} | 
 \simeq 0.99 \times 10^{11} ~E_{\rm TeV} ~ ~ {\rm cm}.
\end{eqnarray*}
These  lengths  are long with respect to
the Earth's  radius 
($R_\oplus \simeq 6.371 \times 10^{8}$~cm), but are 
very short 
 with respect to the typical size
of  astrophysical  sources. Therefore
oscillations are negligible  for  atmospheric  $\nu$  above
$E_\nu \simeq 1$~TeV,  but can be safely averaged
for  (essentially all)  astrophysical neutrinos.
After  space  averaging, the 
oscillation probability matrix    can  be written as:
\begin{eqnarray}
\langle P (\nu_\alpha \to \nu_\beta) \rangle & = & 
\langle P (\overline{\nu}_\alpha \to \overline{\nu}_\beta) \rangle 
 =  
\sum_j | U_{\alpha j}|^2 \,  | U_{\beta j}|^2
\nonumber \\
& \simeq  &
\pmatrix {
0.6  & 0.2  & 0.2 \cr
0.2  & 0.4  & 0.4 \cr
0.2  & 0.4  & 0.4 }
\label{eq:probosc-ave}
\end{eqnarray}
where we have  used the best fit  choice for
the mixing  parameters 
($\theta_{23} = 45^\circ$, 
$\theta_{12} \simeq 32.3^\circ$, 
$\theta_{13}=0$).
The most important  consequence of  (\ref{eq:probosc-ave})
is  a  robust prediction,
valid  for  essentially all   astrophysical
sources,  for the  flavor  composition of the 
observable $\nu$  signal:
$$
\{
\nu_e + \overline{\nu}_e ,
\nu_\mu + \overline{\nu}_\mu ,
\nu_\tau + \overline{\nu}_\tau
\}
= 
\{1,1,1\}
$$

\section{The Gamma--ray sky}
The TeV energy range   (the highest avaliable)  for photons
is the most interesting  one  for  neutrino astronomy, because
the  $\nu$ signal  for the  future telescopes
is expected to be   dominated   by neutrinos  of 
one (or a few) order(s) of magnitude  higher energy.
Recently the  new  Cherenkov   $\gamma$--ray telescopes have
obtained  remarkable results, and the catalogue  of high  energy gamma
ray  sources has  dramatically   increased. Of particular importance
has  been the  scan of the galactic  plane performed by the 
HESS telescope \cite{hess-scan,hess-scan1}, because  for the first time
a  crucially important  region of the sky has  been 
observed  with an approximately   uniform sensitivity
with TeV photons.

The  three brightest galactic  TeV sources
detected   by the  HESS   telescope
 have  integrated  fluxes   above 1~TeV
(in units of $10^{-11}$~(cm$^2$~s)$^{-1}$)
of approximately   2.1 (CRAB Nebula),
 2.0 (RX J1713.7--3946) and 1.9  (Vela Junior)

The fundamental problem  in the 
interpretation of the $\gamma$--ray  sources is the fact
that it is not known  if the observed photons have
hadronic  ($\pi^\circ$ decay)  or  leptonic 
(inverse Compton scattering  of relativistic  electrons
on radiation fields) origin.
If the leptonic mechanism  
is acting, the hadronic  component
is poorly (or not at all)  constrained, and  
the  $\nu$  emission 
can be   much smaller than the $\gamma$--ray flux.
For the hadronic  mechanism,  
the $\nu$ flux  is  at least as
large as the photon flux, and  higher if 
the  $\gamma$--rays in the source are absorbed
(see sec.~\ref{sec:nuhadgam}).

The $\gamma$--ray TeV sources, belong to 
several different   classes.
The Crab is a Pulsar Wind  Nebula, powered  by the
spin down by the central   neutron  star.
The emission from these objects is commonly
attributed   to leptonic  processes, and in particular the
Crab is well  described by the  Self Synchrotron  Comptons
model (SSC). 
The next two   brightest  sources  \cite{hess-velajunior,hess-RX,hess-RX1}
are  young SuperNova Remnants (SNR),   and  there are
good reasons  to  believe  that the photon emission
from these objects  is of hadronic  origin.
The   $\gamma$   spectra  are  power law with a slope
$\alpha \simeq 2.0$--2.2  which  is  consistent
with the   expectation of  the spectra of hadrons
accelerated with   1st order Fermi mechanism   by the 
SN  blast wave. 
The extrapolation  from the photon to the neutrino 
flux  is  then  robust, the main uncertainty being  the possible
presence of a high energy cutoff in the spectrum.

Other TeV  sources that are very promising for
$\nu$ astronomy  are the Galactic  Center \cite{hess-gc}  with
a  measured flux $\sim 0.2$
(same units: $10^{-11}$~(cm$^2$~s)$^{-1}$)
 and 
the micro--Quasar LS5039 \cite{hess-ls5039},
the weakest observed TeV source with a flux  $\sim 0.12$.
These sources  are not particularly   bright in  TeV
$\gamma$--rays,  but there are reasons  to  believe that 
they could  have significant internal absorption for
photons, and therefore have strong  $\nu$ emission.

In general, under the hypothesis of 
(i) hadronic  emission and (ii)  negligible $\gamma$ absorption
(that correspond to $\phi_\nu \simeq \phi_\gamma$),
even we  assume  that  the  $\gamma$ and $\nu$ 
spectra extends  as a power law up to 100~TeV or more, 
the HESS  sources are  just at (or below) the  level 
of  sensitivity of  the  new  $\nu$ telescopes
as will be discussed in the next section.

For  extra--galactic  sources  the constraints  from 
TeV  photon  observations are less stringent  because 
very high energy photons are 
severely  absorbed  over extra--galactic distances 
due to  $\gamma\gamma \to e^+e^-$ interactions on the
infrared  photons (from  redshifted starlight) 
that fill intergalactic space.
Active Galactic  Nuclei  (AGN) are strongly  variables 
emitters of high energy radiation.
The EGRET  detector  on the CGRO satellite
has detected 
over 90  AGN of the Blazar class\footnote{
Blazars are  AGN  that emit  one jet  nearly parallel
(within an angle $\Gamma^{-1}$,
 where $\Gamma$ is the  bulk Lorentz factor of the  jet) 
to the line of sight.}
with  $E_\gamma \ge 100$~MeV \cite{egret-blazars}.  
The  brightest AGN  sources in  the
present TeV $\gamma$--ray catalogue, 
are  the   nearby ($z \simeq 0.03$) AGN's
Mkn 421 and 501 that are  strongly variable
on all  the time  scales considered from few  minutes to  years.
Their average  flux  can be,   during some  periods 
of time, several times  the Crab.
The extrapolation to the $\nu$ flux 
has  significant  uncertainties  because the  origin
(hadronic  or leptonic)  of the detected  $\gamma$--rays is 
not established.

Leading  candidates as  
sources of   high energy neutrinos
are also  Gamma Ray Bursts (GRB)
 \cite{Waxman-GRB,Piran-grb,zhang-meszaros,dar-derujula-grb}.
It is possible  that individual 
GRB  emits  neutrino fluences  that  are sufficiently high to
give  detectable  rates in the neutrino telescopes.

\section{Point Source Sensitivity}
\label{sec:point}
\begin{figure}[ht]
 \includegraphics[width=5.2cm,angle=90.]{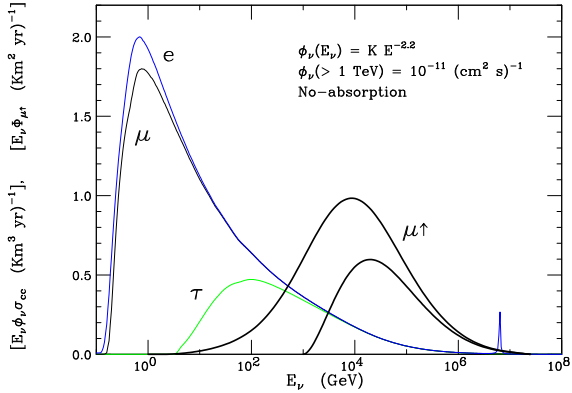}
   \caption{\footnotesize 
Event  rates  from a $\nu$ point source.
The source has 
spectrum $\phi_\nu = K \; E_\nu^{-\alpha}$ with
slope  $2.2$ and normalization 
$\phi_\nu (> 1~{\rm TeV}) = 10^{-11}$~(cm$^2$~s)$^{-1}$
equaly divided  among 6 $\nu$ types.
The thin  lines describe the rate of  $\nu$--interactions in
 the detector volume  for $e$, $\mu$ and $\tau$ like  CC interactions.
The thick lines are fluxes of $\nu$--induced muons 
with 1~GeV  and 1~TeV  threshold.
\label{fig:point1}  }
\end{figure}

The most promising  technique for 
the detection  of  neutrino point sources is 
the   detection of $\nu$--induced  muons ($\mu\uparrow$).
These particles are produced in  the charged current
 interactions of $\nu_\mu$ and  $\nubar_\mu$ in the matter below
the detector.
To illustrate this  important point  we can consider a
``reference'' $\nu$ point source\footnote{
In this  work we have chosen to  characterize the normalization
of a $\nu$ point source  as the flux
(summed over all $\nu$ types)   above a minimum energy
of $E_{\nu, {\rm min}} =1$~TeV.  The reason  for  this  choice
is  that it allows an immediate comparison with the 
the sources  measured  by TeV $\gamma$--ray telescopes,
that are commonly  stated as  flux above $E_{\gamma, {\rm min}} =1$~TeV.
In  case of negligible $\gamma$ absorption
one has $\phi_\gamma \simeq \phi_\nu$.
Since  we consider  power  law fluxes, it is  trivial  to
restate the normalization in other forms.
As  discussed later the km$^3$ telescopes 
sensitivity peaks at 
$E_\nu \sim 20$~TeV.
}
with an  unbroken  power law  spectrum 
of slope $\alpha =2.2$,  and an absolute  normalization
(summing over all $\nu$ types)
$\Phi_\nu (\ge 1~{\rm TeV}) = 10^{-11}$~(cm$^2$~s)$^{-1}$.
The reference source  flux
 corresponds to approximately one half 
the flux  of the   two brightest  SNR 
detected by HESS.
The event  rates  from the  reference  source
of  $\nu$ interactions   with vertex in the
detector  volume, and for $\nu$--induced muons
are shown  in  fig.~\ref{fig:point1},
\begin{figure}[ht]
 \includegraphics[width=5.2cm,angle=90.]{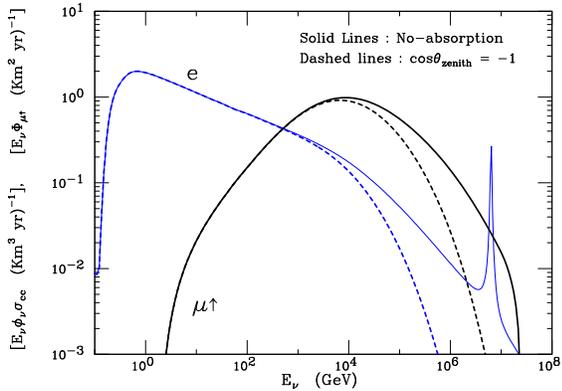}
   \caption{\footnotesize 
Event rates for $e$--like CC   interactions 
(thin lines)  and  $\nu$--induced muons 
with a 1~GeV threshold (thick lines). 
The  source  is the same as  in fig.~\ref{fig:point1}). 
The solid (dashed)  curve are for no (maximum) absorption in the Earth.
\label{fig:point3}  }
\end{figure}
The event rates for $e$, $\mu$ and $\tau$ like 
$\nu$ interactions are:
10.3, 9.6 and 2.9  events/(km$^3$~yr) 
(assuming a water filled  volume),
 while the $\nu$--induced  muon flux
is $\phi_{\mu \uparrow} \simeq 5.6$~(km$^2$~yr)$^{-1}$.
The  signal depends on the  zenith angle
of the source, because of  $\nu$ absorption in the Earth
as  shown in fig.~\ref{fig:point3}.
The  rate of $e$--like events has a small contribution 
from the  ``Glashow resonance''  (the process
$\nubar_e  + e^- \to W^-$), visible as
 a peak at $E_\nu \simeq m_W^2/(2 m_e) \sim 6 \times 10^6$~GeV.
Neglecting the Earth absorption the resonance  contributes
a  small rate  $\simeq 0.07$~(km$^3$~yr)$^{-1}$ to the source signal,
however this  contribution 
quickly disappears  when 
the source  drops much below the horizon, because
of  absorption in the Earth.

The  $\mu\uparrow$  rate is much  easier
to measure and  to disentangle 
from the  atmospheric $\nu$ foreground.
A   crucial   advantage   is that  the 
detected  muon   allows a  high precision 
reconstruction of the $\nu$ direction,
because   the   angle $\theta_{\mu\nu}$ is small.

\begin{figure}[ht]
 \includegraphics[width=5.2cm,angle=90.]{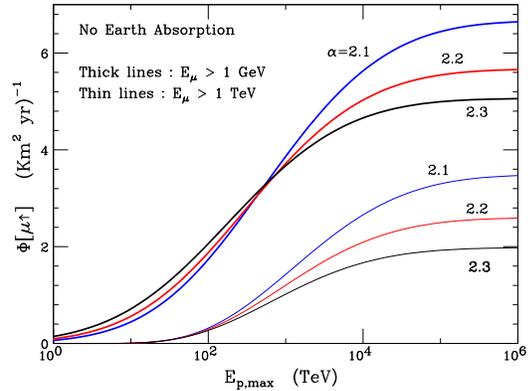}
   \caption{\footnotesize 
Flux of $\nu$--induced  muons,  calculated  for
a  $\nu$ point source   with a power law  spectrum
of slope $\alpha$, plotted as  a function of 
the  high energy cutoff of the parent proton spectrum.
The thick (thin) lines are calculated for 
a threshold $E_{\mu{\rm min}} = 1$~GeV (1~TeV). 
The absolute normalization is  chosen so that
in the absence of cutoff the total $\nu$  flux
above 1~TeV  has  a flux $\Phi_\nu = 10^{-11}$~(cm$^2$~s)$^{-1}$.
\label{fig:signal_mu}  }
\end{figure}

In general the  relation  between the neutrino   and the 
$\nu$--induced  muon  fluxes  is of  order:
\begin{equation}
\Phi_{\mu \uparrow} \simeq   (1 \div 5)
\left [ {\phi_\nu (\ge 1~{\rm TeV} ) \over 
10^{-11} ~({\rm cm}^2 \, {\rm s})^{-1}  } \right ]  
~ ({\rm km}^2 \; {\rm yr})^{-1} 
\end{equation}
The exact  relation
(shown\footnote{
Our  numercal results are in  excellent 
quantitative agreement with the calculation
of Costantini and  Vissani \cite{costantini}  for the case $\alpha=2.2$,
when we  consider  a spectrum with no--cutoff.
The calculation in \cite{costantini}  underestimates 
the  effect of a  cutoff   in the  proton  parent spectrum.
}
in fig.~\ref{fig:signal_mu})
depends on  (i) the slope
of the  $\nu$ spectrum,  (ii) the presence
of  a  high energy cutoff,
(iii)  the threshold  energy used  for
$\mu$  detection  and (iv) (more weakly) on
the  zenith angle of  the source
(because of absorption effects).

The peak of the 
``response'' curve for   the $\nu$--induced  muons
(see fig.~\ref{fig:point1}) is at  $E_\nu \simeq 10$~TeV
(20~TeV  for  a threshold of 1~TeV for the muons),
with a  total  width  extending approximately two  orders of magnitude.
In other words the planned  $\nu$  telescopes  should  be understood
mostly as  telescopes for $\simeq 10$--100~TeV   neutrino  sources.
This  energy range is   reasonably well ``connected''
to  the observations
of  the atmospheric  Cherenkov $\gamma$-ray   
observations that cover the 0.1--10~TeV range in $E_\gamma$, 
and the $\nu$   sources observable in the  planned
km$^3$   telescopes are  likely to be 
appear as bright objects  for 
TeV $\gamma$--ray instruments.

\begin{figure}[ht]
 \includegraphics[width=5.2cm,angle=90.]{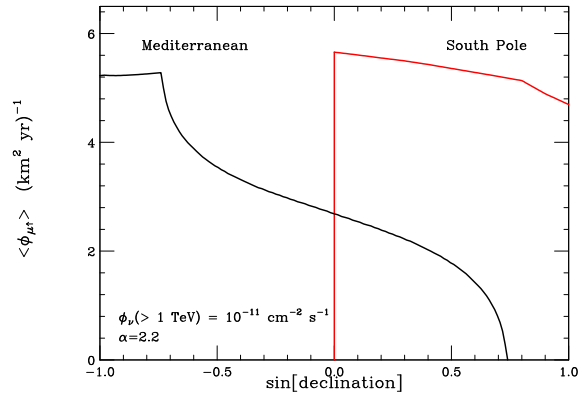}
   \caption{\footnotesize 
Average flux  of up--going  muons  from a $\nu$ point  source
of fixed  luminosity,  plotted
as a  function of  declination  for  two  $\nu$ telescopes
placed  at the south pole (thick line) and in the mediterranean sea
(thin line). The flux is
considered detectable only when the source is below the horizon.
The effects of the (variable)  absorption in the Earth
 is taken into account.
\label{fig:signal2}  }
\end{figure}

Because of the  background  of atmospheric  muons,
the $\nu$--induced  $\mu$'s   are   only  detectable  when the
source is  below  the  horizon.  This
reduces the sky coverage   of a telescope as  illustrated
in fig.~\ref{fig:signal2}, that  gives 
 the  time  averaged  signal   obtained 
by two  detectors,
 placed at the south--pole and in the Mediterranean sea,
when the reference source we are
discussing is placed  at different   celestial   declinations.
For a   south pole detector  the  source  remains at a fixed
zenith angle $\cos \theta_{\rm zenith} = -\sin\delta$, 
and the  declination dependence of the
rate  is  only caused by difference  
in $\nu$ absorption in the Earth for different zenith angles.
The other curves  includes  the effect of the raising
and setting of the source  below the horizon.

Some  of the most promising  galactic  sources
are only visible for a  neutrino telescope in the northern Hemisphere
since  the Galactic Center is  at declination 
$\delta \simeq -29^\circ$.
The  interesting   source
RX J1713.7--3946   is at $\delta = -39^\circ$,
Vela Junior   at  $\delta \simeq -46^\circ$.

\subsection{Background Estimates}
Since the prediction of the 
 signal size for   from the (expected) brightest
$\nu$ sources is  of  only few  events per year,
it is  clearly essential  to  reduce  all
sources of  background to a very small level.
This  is  a possible but remarkably difficult task.

\begin{figure}[ht]
 \includegraphics[width=5.2cm,angle=90.]{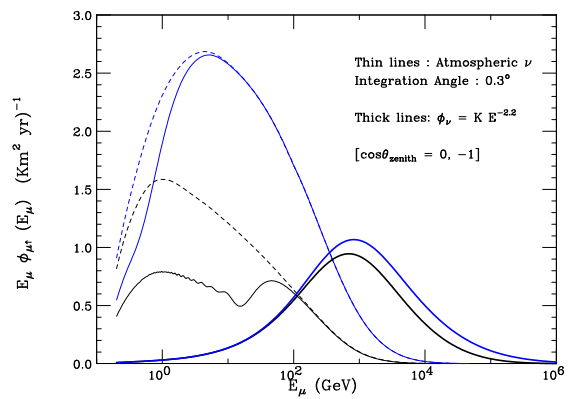}
   \caption{\footnotesize 
Energy Spectra of $\nu$--induced  muons.  The thin lines are the
atmospheric  $\nu$  background integrated  in a cone
of semi--angle $0.3^\circ$
 (dashed  lines   do not  consider
oscillations).  The thick  lines are for a point source
($\alpha =2.2$, $\phi_\nu (> 1~{\rm TeV}) = 10^{-11}$~(cm$^2$~s)$^{-1}$.
The high (low)  line  is for
horizontal (vertical)  muons.
\label{fig:comp_mu}  }
\end{figure}

The  background   problem is  illustrated in fig.~\ref{fig:comp_mu}
that shows the  energy spectrum
 of  the muon signal from our  
``reference''  point  source, 
comparing it  with
the spectrum  of 
the atmospheric  $\nu$  foreground
integrated in a small cone  of semi--angle 0.3$^\circ$.
The crucial  point is that the 
energy  spectrum from  astrophysical  sources
is harder than  the atmospheric  $\nu$ one,  with a median energy
of approximately 1~TeV, an  order of magnitude  higher.
It is for  this reason  that  
for the detection of astrophysical  neutrinos  it is planned
to   use an ``offline''  threshold  of $E_\mu \aprge  1$~TeV.
The atmospheric background    above this threshold
is small  but still  potentially  dangerous,
it depends on the zenith angle and is 
maximum   (minimum) for  the 
horizontal (vertical) direction at the level of 4
(1)~$\mu$/(km$^2$~yr).

The  angular window  for
the  integration of the  muon signal
is  determined by three  factors:
(i) the angular shape of the  source, 
(ii) the  intrinsic angle $\theta_{\mu\nu}$,  and
(iii) the angular   resolution of the instrument.
The  source  dimension can be important 
for  the galactic  sources, in fact 
the  TeV $\gamma$--ray  sources  have a finite size,
in  particular 
the SNR RX J1713.7--3946   has a radius  $\sim 1^\circ$,
and Vela Junior is  twice as large.
The    detailed
morphology   of these sources 
measured by the HESS  telescope  indicates  that 
most of the emission is coming  from  only some  parts
of the shell  (presumably where the gas   density is higher),
and the   detection of   $\nu$   emission  from  these 
sources could  require  the careful  selection   of  
the angular region of the $\gamma$--ray  signal.

\begin{figure}[ht]
 \includegraphics[width=5.2cm,angle=90.]{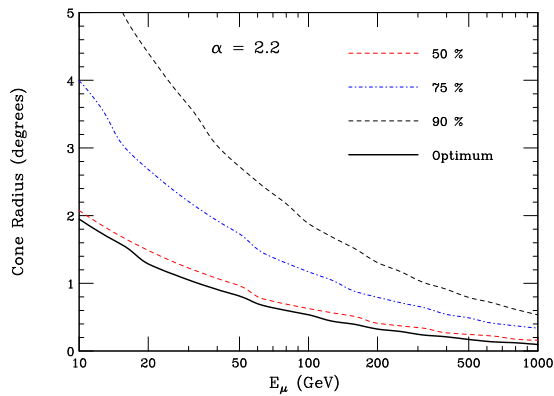}
   \caption{\footnotesize 
Angular distribution of the  upgoing  muon  flux induced
by a $\nu$  power law flux of slope $\alpha = 2.2$.
The curves  give the semi--angle angle 
of the cone
that (for a fixed  value of $E_\mu$) contains
50\%, 75\% and 90\% of the signal.  The thick line
gives the cone  that  maximize  the $S/\sqrt{B}$.
\label{fig:ang0}  }
\end{figure}

The distribution of the  $\theta_{\mu\nu}$ angle 
defines the  minimum  angular dimension of a perfect point source
signal. This is  shown in  fig.~\ref{fig:ang0}
that plots  the  semi--angle  of the cone
that contains 50, 75 or 90\% of such a signal 
(for a $\alpha = 2.2$ spectrum)
as  a function of the muon energy.
The  size the 
muon signal  shrinks with increasing  energy 
$\propto (E_\mu)^{-1/2}$.  This  is easily understood
noting that the  dominant 
contribution to $\theta_{\mu\nu}$  is   the
muon--neutrino scattering angle at the
$\nu$--interaction point: 
\begin{equation}
\cos \theta_{\mu\nu} = 1  - \frac{m_N \,x}{E_{\mu,0}} ~\left (
1 - \frac{E_{\mu,0}}{E_\nu} 
\right )
\end{equation}
($E_{\mu,0}$ is the muon energy at the interaction point),
and expanding for small angle one finds
$\theta_{\mu\nu}  
 \sim \sqrt{m_N \,  x /(2 \,  E_{\mu,0})}$.
The  50\%  containement cone  angle   
shrinks to  $\theta_{50\%} \simeq 0.15^\circ$
for  $E_\mu = 1$~TeV that is  probably smaller
than the experimental resolution.

Qualitatively the experimental strategy for the maximization of the
point--source sensitivity is clear: (i) selection of high energy
$E_\mu \aprge 1$~TeV muons, and (ii) selection of the narrowest
angular window compatible with the experimental resolution and the
signal angular size.
These cuts  will reduce the total
signal by at least a factor of 2.
The optimization of the cuts for signal
extraction is a non--trivial problem,  that  we cannot
discuss here.  
In principle one would like to maximize the quantity $S/\sqrt{B}$ 
(where $S$ and $B$ are respectively
the signal and background event numbers).  For the angular window
(where $B \propto \theta$) this problem has a well defined solution
that is  shown in fig.~\ref{fig:ang0},
that corresponds  to the choice of a window
that contains  approximately 50\% of the signal.
This  is the numerically obtained
generalization of the well known fact that for a gaussian angular
resolution of width $\sigma_\theta$, $S/\sqrt{B}$ 
is maximized choosing the
angular window $\theta \le 1.585 \, \sigma_\theta$   that contains
a fraction 0.715  of the signal. 
For the energy cut,  the quantity
$S/\sqrt{B}$   grows    monotonically when the 
threshold energy is increased,  and therefore
the optimum   is  determined  by the condition   that the 
background   approaches  zero.
 
\subsection{Sensitivity and Sources}
The conclusion of this   discussion on the point--source  sensitivity
is  that   after  careful    work  for     background reduction,
the   minimum    $\nu$   flux  (above 1~TeV) 
detectable as a point source  for  the km$^3$ telescopes
is of   order (or  larger than) 
$\Phi_\nu \sim  10^{-11}$~(cm$^2$~s)$^{-1}$, 
that is  approximately the $\nu$ flux of the 
reference  source  considered above.
This  is  tantalizingly (and frustratingly)  close to  the 
flux    predicted from  the extrapolation
(performed  under the assumption of a transparent 
source)  of the brightest sources in the TeV $\gamma$--ray catalogue.

The  remarkable recent results of HESS and
MAGIC on galactic  sources are certainly  very exciting
for the $\nu$ telescope  builders,
because they show the  richness of the ``high energy'' sky,
but they are also  sobering, because  they lead to a prediction
of the scale   of the brightest  galactic $\nu$ sources, 
that is just at  the
limit of the sensitivity of the planned instruments.
As  discussed  previously,  strong  $\nu$  sources
could lack  a  bright  $\gamma$--ray counterparts
because of internal  photon absorption, or because they
are at extra--galactic distances.

The considerations  outlined in this section
can  be immediately applied to a 
burst--like  transient sources  like GRB's.
In the presence of a  photon ``trigger''
that allows a time  coincidence, 
the  background problem disappears and
the detectability of one  burst is   only defined
by  the signal size.
A burst with an $\nu$ energy fluence 
$F_\nu \simeq 2 \times 10^{-4}$~erg/cm$^2$
per decade of energy  (in the  10--100~TeV range)
will produce an average  of one  $\nu$--induced  muon in
a km$^3$ telescope. 

\section{Extragalactic  Neutrino Sources}
\label{sec:extra}
The flux from the ensemble of all extragalactic  sources,
with the exception of  a few sources will appear
as an unresolved isotropic flux,  characterized  by  its energy spectrum
$\phi(E)$.  
The identification of this unresolved component can  rely on
three signatures: (i) an energy spectrum harder that the  atmospheric
flux, (ii) an isotropic  angular  distribution, and (iii) approximately
equal fluxes for all 6 neutrino types.
This last point is a consequence of space averaged flavor oscillations.

The  flux of  prompt (charm decay) atmospheric  neutrinos
is also approximately isotropic in the energy range considered, 
and  is characterized by equal fluxes of $\nu_e$ and $\nu_\mu$,
 with however
a significant smaller flux of $\nu_\tau$  (that are only produced
in the decay of $D_s^\pm$).  
The disentangling of the  astrophysical  and  an prompt--atmospheric
fluxes is therefore  not trivial, and depends on a  good determination
of the neutrino energy spectrum, together with a  convincing model
for  charmed particle production in  hadronic interactions.
A model  independent method for  the identification of the astrophysical
flux  requires the separate measurement  of the fluxes of all
three  neutrino  flavors, including the $\nu_\tau$.
The flux of prompt atmospheric  $\nu_\mu$ is also  accompanied
by an approximately equal flux  of $\mu^\mp$
(the differences  at the level of 10\%  are due to the very well
understood  differences in the spectra produced in weak decays).
The down--going muon flux is in principle  measurable, determining
the prompt  component experimentally and eliminating a potentially dangerous 
background.  This  measurement is therefore  important and efforts should be
made to  perform it.

\subsection{Energetics of Extra--galactic neutrinos}
The  most  transparent way to   discuss the 
$\nu$ extragalactic  flux
is  to consider its energetics.
The observable $\nu$   flux  $\phi(E)$   is  clearly  related
to  a number density  $n(E) = (4 \pi/c)  \, \phi(E)$ and to an energy
density $\rho(E)  = E \, n(E)$, that  fill uniformly the  universe.
The $\nu$   energy density  at the present  epoch
has  been generated by
a  power  source  acting   during the history of the universe.
Integrating over  all $E_\nu$,  the relation between the
 power  injection   density $\ldens (t)$   and the 
$\nu$  energy density at the present  epoch is given by:
\begin{eqnarray}
\rho_\nu & = &  \int_0^{t_0}  \; dt~ { \ldens (t) \over [1 + z(t)] }  ~ = ~ 
\int_0^{\infty}  \; dz~  \left | { dt \over dz } \right |
~ { \ldens (z) \over (1 + z) }  \nonumber \\
 & ~ & ~ \nonumber  \\
 & = &  \int_0^{\infty}  \; dz~  
{ \ldens (z) \over H(z)  \; (1+z)^2 }
~ = ~ { \ldens_0 \over H_0 } ~ \xi
\label{eq:energy-balance}
\end{eqnarray}
where $t_0$ is the age of the unverse, 
$\ldens_0$  and $H_0$ are the $\nu$ power density 
and the  Hubble constant at the present epoch, and $\xi$ 
 is  the  adimensional
quantity:
\begin{equation}
\xi  = \int_0^\infty \, dz ~
\left [ \frac{H_0}{H(z)} \right ]~
\left [\frac{\ldens(z)}{\ldens_0}\right ] \;(1+z)^{-2} ~~.
\label{eq:xi-def}
\end{equation}
that  depends on the cosmological parameters,
 and more crucially on  the  cosmic
history of the neutrino  injection.
The quantity  $\ldens_0 \, \xi$ can be  understood as the 
effective ``average  power  density''
operating  during  the  Hubble time $(H_0)^{-1}$,
and  depends on the cosmological parameters,
and more strongly on the cosmological  history of the
injection.
For  a Einstein--De Sitter  universe ($\Omega_{\rm m} = 1$, 
$\Omega_{\Lambda} = 0$)  with no evolution for the sources
one has  $\xi = 0.4$, for the ``concordance model'' cosmology
($\Omega_{\rm m} = 0.3$, $\Omega_{\Lambda} = 0.7$)
this becomes $\xi = 0.53$.
For the  same concordance model cosmology, 
if it is assumed  that the cosmic time  dependence of the
neutrino injection is  similar   to the one 
fitted to the star--formation
history   \cite{Heavens:2004sr,fukugita-peebles},
one obtains $\xi({\rm SFR}) \simeq 3.0$,
for a time  dependence equal to the
one  fitted to the AGN  luminosity evolution  \cite{ueda}
one finds  $\xi({\rm AGN}) \simeq 2.2$.

If we  consider the
power and  energy density
not bolometric, 
but  only integrated
above  the  threshold energy  $E_{\rm min}$, 
the redshift effects depend on  the shape of the injection spectrum.
For  a power law   of slope $\alpha$ independent
from cosmic time, 
one has still a  relation 
of form:
\begin{equation}
\rho_\nu(E_{\rm min}) 
= \frac{\ldens_0(E_{\rm min})}{H_0} ~\xi_\alpha
\label{eq:energy-balance1}
\end{equation}
very similar to (\ref{eq:energy-balance}), with: 
\begin{equation}
\xi_\alpha  = \int_0^\infty \, dz ~
\left [ \frac{H_0}{H(z)} \right ]~
\left [\frac{\ldens(z)}{\ldens_0}\right ] \;(1+z)^{-\alpha}
\label{eq:xi-def1}
\end{equation}
For   $\alpha=2$ one has $\xi_2 = \xi$.
The energy density (\ref{eq:energy-balance1})  corresponds to
the $\nu$ isotropic  differential   flux:
\begin{eqnarray}
\phi_\nu (E)  & = & 
 K_\nu ~E^{-\alpha} \nonumber  
\nonumber \\
&  = &    \left (\frac{\mathcal{L}_0 (E_{\rm min})
 \, E_{\rm min}^{\alpha-2} \, \xi_\alpha \, k_\alpha}
{4 \pi \, H_0} \right ) ~ E^{-\alpha}
\label{eq:flux-extra}
\end{eqnarray}
where $k_\alpha$ is the kinematic adimensional factor:
\begin{equation}
k_\alpha \simeq (\alpha-2) /
[1 -  (
 E_{\rm min}/E_{\rm max} )^{\alpha-2} 
]^{-1}
\label{eq:kalpha}
\end{equation}
with $E_{\rm max}$  the high energy cutoff of the  spectrum.
The limit of $k_\alpha$ when   $\alpha \to 2$ is: 
$k_\alpha =  \left [\log (E_{\rm max}/E_{\rm min}) \right]^{-1}$.
One can use  equation (\ref{eq:flux-extra}) to relate
the observed diffuse  $\nu$   flux 
to the   (time and space) averaged   power  density
of the neutrino  sources:
\begin{equation}
(K_\nu \; E_{\rm min}^{2 -\alpha}) = 
 \frac{\ldens_0 (E_{\rm min})
 ~\xi_\alpha ~ k_\alpha}{ 4 \pi\; H_0}
\label{eq:K-energy}
\end{equation}
The case $\alpha=2$  corresponds to an equal power
emitted  per  energy decade, and  numerically:
\begin{equation}
K_\nu \simeq 3.7 \times 10^{-11} ~
\left [ 
\frac{(\ldens_0  \, \xi)_{\rm decade}}{L_\odot/{\rm Mpc}^3}
\right ]
~\frac{\rm GeV}{{\rm cm}~{\rm s}~{\rm sr}}
\label{eq:K-energy-2}
\end{equation}
($L_\odot$ is the solar luminosity).
The current  limit on the existence
of a diffuse, isotropic $\nu$ flux
of slope  $\alpha \simeq 2$
from the AMANDA and Baikal detectors 
corresponds to $K_\nu \aprle 8 \times 10^{-7}$~GeV/(cm$^2$~s~sr)$^{-1}$,
while the    sensitivity of  the future
km$^3$ telescopes  is   estimated  (in the same  units) 
of order 
$K_\nu \sim 3 \times 10^{-9}$. 
This  implies  that the 
average power density  for the  creation of high energy neutrinos
is: 
\begin{equation}
(\ldens_0 \xi) \aprle 
2.4 \times 10^4~\lum0~{\rm decade}^{-1}
\end{equation}
the future km$^3$ telescopes  will detect the  extragalactic  flux
if it has been  generated  with an ``average''
power  density   $ (\ldens_0 \, \xi) \aprge 80$
(same units).

\subsection{Possible Sources of Power}
To evaluate the astrophysical  significance 
of the  current  limits on the  diffuse
$\nu$ flux  and of the expected  sensitivity of
the km$^3$  telescopes, one can  consider 
the energetics of the most plausible sources of high
energy neutrinos,  namely the death of massive stars
and the activity around  Super Massive Black
Holes at the center of galaxies.
\subsubsection{The Death of Stars}
The bolometric  average  power  density
associated  with stellar  light  has  been  estimated 
by Hauser and Dwek \cite{hauser-dwek} as:
\begin{equation}
(\ldens  \,\xi_{\rm star})^{\rm bol}
= (0.36 \div 1.23) \times 10^9
  \lum0 
\end{equation}
this is also in  good  agreement with the  estimate of
Fukugita  and Peebles
\cite{fukugita-peebles}
for the B--band optical luminosity
of stars.
This power  implies   that the  total
 mass density gone into the star formation 
\cite{fukugita-peebles} is:
\begin{equation}
\rho_{\rm star} \simeq 5.9 \times 10^8 ~
\left (
{ M_\odot ~{\rm Mpc}^{-3} }
\right )
\end{equation}
(that  implies 
$\Omega_{\rm star} = 0.0043 \simeq 0.09 \; \Omega_{\rm baryon}$).
Most of this power has  of course little to
do  with cosmic rays and high energy  $\nu$  production,  however  
c.r.  acceleration has  
been associated with the   final  stages of 
the massive stars life.
A well established  mechanism  is acceleration 
in the spherical  blast waves  of SuperNova explosions, a more
speculative one  is acceleration in   relativistic jets  that are
emitted in all or a subclass  of the SN, and that  
is the phenomenon
behind the (long duration) Gamma Ray Bursts.
The current estimate  of the  rate of gravitational collapse 
supernova rate   obtained  averaging  over different
surveys  \cite{fukugita-peebles} is:
\begin{equation}
R_{\rm SN}^{\rm observed}  
\simeq  7.6^{+6.4}_{-2.0}   \times  10^{-4} 
~ ({\rm Mpc}^3 ~{\rm yr})^{-1}
\end{equation}
This  is  reasonably  consistent with the 
the estimate of  the power  emitted   by  stars, 
if one assumes  that  
all  stars  with $M > 8~M_\odot$  end their history 
with a gravitational collapse.
Fukugita and Peebles  
estimate ($\psi(0)$ is the SFR at thre present epoch):
\begin{eqnarray}
R_{\rm SN} 
& = & \psi(0) ~ { \int_8^{100}~ dM ~ {dN \over dM}   \over
 \int_{0.08}^{100}~ dM ~  M~ {dN \over dM} } 
\nonumber \\
&\simeq   & 7.9^{+2.4}_{-3.9}   \times  
10^{-4} ~ ({\rm Mpc}^3 ~{\rm yr})^{-1} 
\end{eqnarray}
Assuming    that  the  average kinetic energy
released in a SN  explosion is of order
$\langle E_{\rm kin} \rangle_{SN} \simeq 1.6 \times 10^{51}~{\rm erg}$
this  implies  an average    power  density: 
\begin{equation}
\left ( \ldens  \; \xi \right )_{{\rm SN},{\rm kin}}
  \simeq 
4.2 \times 10^{6}  ~
\lum0 
\end{equation}
It is  commonly believed that a fraction of order 
$\sim 10$\% of this  kinetic  energy is  converted  into
relativistic hadrons. One can see that if on average  a fraction
of order 1\% is  converted in neutrinos,   this  could  result  in 
$(\ldens_0 \, \xi)_\nu \sim 10^{3}~\lum0$, that
could  give a  a  detectable signal from 
the ensemble of  ordinary galaxies.

\subsubsection{Active Galactic Nuclei}
Active  Galactic  Nuclei  have often been suggested as a  source
of   Ultra High energy Cosmic  Rays, and at the same  time
of  high energy neutrinos.
Estimates of the  present epoch  power density 
from  AGN in the 
[2,10~KeV] $X$--ray  band are of order
\begin{equation}
\ldens_{0,{\rm AGN}}^{[2,10\,{\rm KeV}]}   \simeq  2.0 \times 10^{5}  ~
\lum0
 \end{equation}
with a  cosmic  evolution \cite{ueda}  characterized
by $\xi \sim 2.2$.
Ueda \cite{ueda}  also estimates the 
relation between this energy band  and the 
bolometric  luminosity is:
$
L_{\rm bol} \simeq 30~L_{X}^{[2,10\,{\rm KeV}]}.
$
and therefore
the total  power associated with AGN is 
\begin{equation}
\left ( \ldens \, \xi \right )_{\rm AGN}^{\rm bol}   \simeq  10^{7}
  ~ \lum0
\label{eq:lumagn}
 \end{equation}
of order  $\sim 1$\% of the power associated with star light.
It is  remarkable  that this AGN  luminosity is
well matched to the average mass  density
in Super Massive Black Holes, that is
estimated  \cite{marconi,ferrarese-ford}:
\begin{equation}
\rho_\bullet = 
4.6^{+1.9}_{-1.4} \times 10^5 ~M_\odot ~{\rm Mpc}^{-3}
\label{eq:rhobh}
\end{equation}
When a  mass  $m$  falls into a Black Hole (BH) of mass $M_{\bullet} $
the BH  mass  is  increased by an amount: 
$\Delta M_{\bullet} =  (1 - \varepsilon) \, m $
while the  energy $\varepsilon \, m$ is  radiated away 
in different  forms.
The total accreted   mass can therefore be
related to the amount of radiated  energy, if  one knows
the  radiation efficiency $\varepsilon$:
\begin{equation}
M_{\bullet} = {(1 -\varepsilon) \over \varepsilon } ~ E_{\rm radiated} 
\end{equation}
and  therefore
\begin{equation}
\rho_\bullet = (\ldens \, \xi)_{\rm AGN} ~ H_0^{-1}
  \frac{1-\varepsilon}{\varepsilon}
\end{equation}
The estimates (\ref{eq:lumagn})
and  (\ref{eq:rhobh}) are consistent for an  efficiency $\varepsilon  
\simeq 0.1$. 
Assuming  radiated  energy is the 
kinetic energy accumulated in the fall  down
to a  radius  $r$  that is $f$ times the BH Schwarzschild
radius  $R_S = 2 G \, M_\bullet$:
\begin{equation}
\varepsilon = {E_{\rm radiated}  \over m}
\simeq { G \, M_{\bullet} \over f \, R_S}  
= {1 \over 2 \, f }  
\end{equation}
this  corresponds to   $f\sim 5$.

The ensemble of  AGN 
has sufficient power to generate a diffuse $\nu$ flux
at the level of the existing  limit.

\subsection{Gamma Ray Bounds}
A  very  important guide for the  expected flux of
extragalactic   $\nu$  is the diffuse
$\gamma$--ray flux  measured  above  100~MeV  measured 
by the EGRET  instrument \cite{egret-extra} that can be 
described
(integral  flux  above a minimum energy $E_\gamma$ in GeV)
as:
\begin{equation}
\Phi_\gamma (E_\gamma) \simeq 
1.42 \times 10^{-6} 
E_\gamma^{-1.1}
({\rm cm}^2\,{\rm s} \, {\rm sr})^{-1}
\label{eq:egret-diffuse}
\end{equation}
(for a critical  view  of this
interpretation see \cite{Strong:2005zn}).
This  corresponds to the energy density 
in the decade between  $E_\gamma$ and $10\, E_\gamma$
(with $E_\gamma$ in GeV):
$\rho_\gamma (E_\gamma)  \simeq 
1.35 \times 10^{-6} ~E_{\rm min} ^{-0.1}
\,  {\rm eV}~{\rm cm}^{-3} 
$
and to the average power density per decade:
\begin{equation}
\left( \ldens  \; \xi \right )_{\gamma}
  \simeq 
3.7 \times 10^{4} 
~ {E_\gamma}^{-0.1}
~ \left ( 
L_\odot ~{\rm Mpc}^{-3} \right ) 
\end{equation}
This  diffuse $\gamma$--ray 
has  been attributed \cite{stecker-salamon}  to the emission  of unresolved 
Blazars.
The sum of the measured fluxes  (for $E_\gamma  > 100$~MeV) 
for all identified  EGRET blazars  \cite{egret-blazars}
amounts to:
$ \sum_{{\rm Blazars}}  \phi_j  \simeq 0.33 ~({\rm m}^2 \,{\rm s})^{-1}$
This  has to  be compared with  the   diffuse flux 
(\ref{eq:egret-diffuse})
that can   be  re-expressed as:
$ 4 \pi \;   \Phi_\gamma (100~{\rm MeV}) 
 \simeq 2.62 ~({\rm m}^2 \, {\rm s})^{-1}$.
Other authors 
\cite{mukherjee} arrive to the
interesting conclusion 
that  unresolved  blazars can account  
for  at most 25\% of the diffuse flux,  and that  there  must be 
additional  sources of GeV photons.

Most models for  
the Blazar emission  favor  leptonic mechanism,  with the  photons
emitted by the Inverse Compton scattering  of relativistic 
electrons  with radiation fields   present  in  the jet, however
it is possible that   hadrons
account for a  significant fraction (or even most) 
of the $\gamma$--ray flux, and therefore that the  $\nu$ flux
is of  comparable intensity to the $\gamma$ flux.
The  test of  the ``Egret level'' for the diffuse
$\nu$ flux  is one of the  most important  goals for the
neutrino  telescopes already in operation.

\subsection{Cosmic Rays Bounds}
Waxman and Bahcall \cite{upper-bound}
have suggested the existence of an upper bound
for the diffuse flux of extra--galactic  neutrinos 
that is   valid if the $\nu$ sources  are   transparent
for  cosmic rays, based on the observed  flux of ultra
high energy cosmic rays.
The logic (and the limits) of the WB bound  are  simple to grasp.
Neutrinos are produced  by cosmic  ray interactions, and
it is  very likely (and certainly economic)  to assume that
the c.r.  and $\nu$ sources  are the same, 
and that the  injection rates for the two type of particles
are related.
The condition of ``transparency'' for  the  source
means  that a  c.r.  has  a probability
$P_{\rm int} \aprle  0.5$  of  interacting  in its way out of the source.
The  transparency condition 
obviously  sets an upper bound on the  $\nu$  flux,
that can  be  estimated  from  a  knowledge 
of the cosmic ray  extragalactic  spectrum,
calculating for each  observed  c.r.  the spectrum
of neutrinos  produced in  the shower
generated by the interaction of one  particle of  the same 
mass  and energy, and integrating over  the
c.r.  spectrum.  If the c.r. spectrum is  a power
law  of form $\phi_{\rm c.r.} (E) = K \; E^{-\alpha}$,
the ``upper bound''  $\nu$  flux   obtained  saturating the
transparency condition (for a c.r. flux dominated by protons)  is:
$\phi_\nu = (K \, Z_{p \to \nu}) \; E^{-\alpha}$
with $Z_{p \to \nu} \sim 0.25$ 
 
\begin{figure}[ht]
 \includegraphics[width=5.2cm,angle=90.]{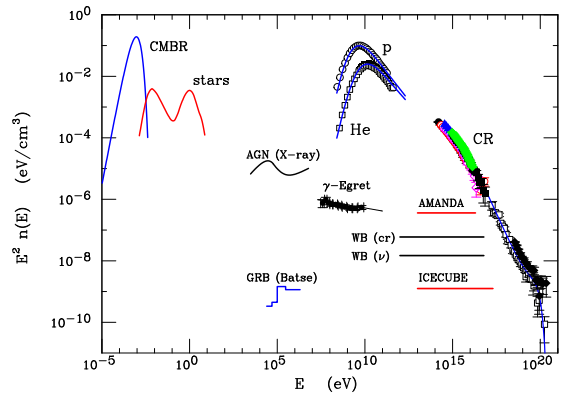}
   \caption{\footnotesize 
Different contributions to the  average energy density  in 
the  universe.  Different lines describe  the contributions of
the 2.7$^\circ$~K  CMBR  radiation, starlight (with reprocession by dust),
the X-ray diffuse flux (attributable to AGN, the  Egret 
($E_\gamma \ge 100$~MeV)  diffuse flux, and the GRB contribution from
the BATSE data.
The detected cosmic  ray   energy density  (points) 
is  enhanced by magnetic  confinement
in our galaxy and is not cosmological, except (possibly) at the 
highest energy. 
The lines labeled WB(cr)  and WB($\nu$) are the
extragalactic c.r. and upper bound $\nu$ fluxes estimated
in \cite{upper-bound}. 
The line  labeled Amanda  is the current limit  on
an isotropic extragalactic $\nu$ flux, and the line labeled  ICECUBE
is the predicted  sensitivity on km$^3$ telescopes.
\label{fig:energy_density}  }
\end{figure}

The WB upper  bound (shown in fig.~\ref{fig:energy_density}) 
has  been the object of several criticisms  see for example
\cite{MPR}. There are two problems with it. 
The first one  is  conceptual:
the condition of transparency is  plausible 
but is not physically  necessary. 
The  c.r. sources can   be 
very transparent,  for example in  SNR's 
the interaction probability of the hadrons accelerated
by the blast  wave, 
is  small ($\ll 1\%$)
(with a value that depends  on the density of
the local ISM),  but
``thick  sources''  are possible, and have  in fact 
been advocated for a long time,
the  best example  is acceleration
in the vicinity of the  horizon of a SMBH.
The  search for ``thick''  neutrino sources
is after all one of the important  motivations
for $\nu$ astronomy. 

The second problem is only quantitative.
In order to estimate the bound one  has to know the spectrum of
extra--galactic  cosmic rays.  
This  flux  is  ``hidden''  behind   the 
foreground of galactic  cosmic rays, that
have a  density enhanced  by 
magnetic confinement   effects.
The separation of the galactic and extragalactic  component 
of the c.r.  is a  central unsolved  problem 
for cosmic ray science. 
The extragalactic  c.r.  is  dominant
and  therefore  visible only 
at the very highest energies, perhaps  only above the
``ankle''  (at $E_{\rm ankle} \simeq 10^{19}$~eV) 
as  assumed  in \cite{upper-bound}.
More recently it has  been argued \cite{Berezinsky:2005cq}
that extra--galactic $p$ dominate the c.r. flux for  $E_0 \aprge
10^{18}$~eV.
Waxman and Bahcall have fitted  the c.r. flux
above the  ankle  with an  a $E^{-2}$ injection spectrum
(correcting for energy degradation in the  CMBR) 
 and extrapolated the flux with the same form
to  lower energy.
The  power needed  to   generate this  c.r. density 
was $\ldens_{\rm cr} \simeq 1900~L_\odot/$(Mpc$^{3}$~decade).
The extrapolation of the c.r.
flux (that is clearly model dependent)
and the corresponding $\nu$ upper bound are shown 
in fig.~\ref{fig:energy_density}, where they 
can be compared to the sensitivity of the $\nu$ telescopes.

Because  of the uncertainties
in the fitting of the  c.r.  extragalactic  component
and its  extrapolation  to low energy
(and the possible loophole of the existence of
thick  souces) 
the WB estimate  cannot really  be considered an
true  upper bound  of the $\nu$ flux.  However 
the ``WB  $\nu$--flux''  is  important as
reasonable (and indeed in many senses  optimistic)
estimate of the  order of magnitude 
of the true  flux.
The existence of cosmic rays
with energy as large as
$E_0 \sim 10^{20}$~eV  is  the best 
motivation for  neutrino   astronomy, 
since  these particles
must be accelerated to  relativistic energies
somewhere  in the universe, and   
{\em  unavoidably}  some  c.r.
will interact with target  near (or in) the source
producing  neutrinos.

\subsection{Resolved  and Unresolved Fluxes}
An interesting question is the relation
between the  intensity of the   total extragalactic  $\nu$ 
contribution, and the potential to  identify extra--galactic
point sources.  
This clearly depends on the luminosity function and cosmological
evolution  of the sources.
Assuming for  simplicity that all sources have 
energy spectra of the same shape, and in particular power spectra
with slope $\alpha$, then  each source can be  fully described  by
its  distance and its $\nu$ 
luminosity $L$ (above a  fixed  energy threshold $E_{\rm min}$).
The ensemble  of all extra--galactic sources is then described
by the   function  $n(L, z) \, dL$  that gives the   number
of  sources 
with luminosity in the interval between 
$L$ and $L + dL$
contained in a unit  of comoving volume 
 at the epoch  corresponding to  redshift $z$.
The power  density due to the ensemble of all sources
is given by:
\begin{equation}
\ldens (z)  = \int dL ~L ~n(L, z)
\label{eq:ldens}
\end{equation}
It is is possible (and in fact  very likely) that most 
high energy neutrino sources 
are  not be isotropic. This  case is  however   contained
in our discussion if  the 
luminosity $L$ is  understood  as an orientation dependent
``isotropic luminosity'': 
$L \equiv (4 \pi) \, (dL_{\rm true}/d\Omega)$.
For a random
distribution  of the viewing angles it is 
easy to show  that
\begin{equation}
\int dL ~L ~n(L, z) = 
\int dL_{\rm true} ~L_{\rm true} ~n(L_{\rm true}, z)
\end{equation}
The $\nu$  flux  (above the threshold  energy
$E_{\rm min}$) received  from a  source 
described by $L$ and $z$ is:
\begin{equation}
\Phi_{\rm point} (L, z) = 
\frac{L}{4 \pi \; E_{\rm min} } ~
\frac{k_\alpha}{(\alpha-1)} ~
\frac{(1+z)^{2-\alpha}}{d_L(z)^2}
\label{eq:phipoint}
\end{equation}
where 
$d_L(z)$ is the luminosity distance 
and $k_\alpha$  is 
the  adimensional  factor given in (\ref{eq:kalpha}).

\begin{figure}[ht]
 \includegraphics[width=5.2cm,angle=90.]{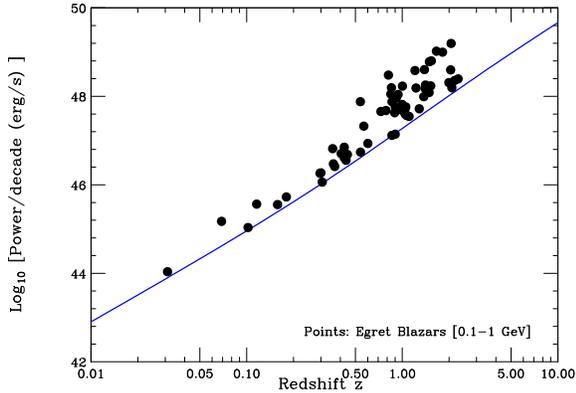}
   \caption{\footnotesize 
Redshift horizon  $z_h(L)$ 
calculated  for the  cosmology
$\Omega_{\rm m} = 0.3$, $\Omega_\Lambda = 0.7$, and
sources with power law spectra of
slope $\alpha =2$,  assuming the a  minimum detectable
flux  $\Phi_\nu (\ge 1~{\rm TeV} = 10^{-11}$~(cm$^2$s)$^{-1}$.
The points corresponds to the EGRET blazars \cite{egret-blazars}
with the luminosity measured in the 0.1--10~GeV range.
\label{fig:horizon}  }
\end{figure}

If $\Phi_{\rm min}$ is the sensitivity of a  neutrino  telescope,
that is the minimum flux  for the detection of a point source,
then a source of luminosity $L$ 
can be detected only if is  closer than a maximum  distance
corresponding to redshift $z_h(L)$. 
Inspecting  equation (\ref{eq:phipoint})
it is  simple to  see that  $z_h$ is a  function of the
adimensional ratio $x =  \sqrt{L/L^*}$ 
where:
\begin{equation}
 L^* = 4 \pi \, E_{\rm min} \, \Phi_{\rm min}~(\alpha-1)/(H_0^2\;k_\alpha)
\label{eq:lstar}
\end{equation}
is the order of magnitude  of  the  luminosity of a source that  gives
the minimum  detectable flux  when placed at $z \sim 1$.
The explicit solution   for $z_h(x)$
depends on the cosmological  parameters ($\Omega_{\rm m}$,
$\Omega_\Lambda$) and on the spectral slope $\alpha$.
A general  closed form  
analytic solution for $z_h(x)$ does not exist\footnote{
As an example, for $\Omega_{\rm m}=1$,
$\Omega_\Lambda=0$ and $\alpha =2$ the exact solution is
$z_h = (x -1 + \sqrt{1+2 \,x})/2$.},
but it  can be easily obtained  numerically.
The solution for the concordance model  cosmology
is shown in fig.~\ref{fig:horizon}.
It is  also  easy and useful  to  write   $z_h$ 
as a power law  expansion in $x$:
\begin{equation}
z_h  = x + \frac{1}{4}
\left [ 2(1 -  \Omega_\Lambda) + \Omega_{\rm m} -2 \alpha 
\right ]\, x^2  + \ldots 
\end{equation}
The leading term of  this  expansion   is   simply $x$ 
independently  from the cosmological parameters  and the slope
$\alpha$. 
This clearly  reflects   the fact that  for small redshift $z$ 
one probes  only the near universe,
where and when redshift effects   and 
cosmological evolution are  negligible,
and the flux  simply scales as the inverse square of the distance.

The total flux  from all  (resolved and unresolved)  sources
can be obtained  
integrating  over $L$ and $z$:
\begin{equation}
\Phi_{\rm tot} =
\frac{1}{4 \pi} \int dL 
\int_0^\infty  dz \;
v(z) \,
n (L, z)
\; \Phi_{\rm point} (L, z)
\end{equation}
where $v(z)$ is the  comoving volume  contained   between 
redshift $z$ and $z+dz$:
\begin{equation}
v (z)  =  4 \pi \; 
~ \frac{d_L^2 (z)}{(1 + z)^2 \; {H}(z)}
\simeq \frac{4\pi}{H_0^2} \, z^2 \, (1 + \ldots)
\label{eq:vol1}
\end{equation}
Substituting the  definitions 
(\ref{eq:phipoint})  and (\ref{eq:vol1}), 
integrating in $L$ and  using  (\ref{eq:ldens}) 
the total flux corresponds exactly to  the
result (\ref{eq:flux-extra}).
The resolved  (unresolved)   flux can be obtained
changing the limits  of integration in  redshift
to the interval  $z \in [0, z_h(L)]$   ($z \in [z_h(L),\infty]$).
The total number of  detectable  sources
is obtained  integrating the source density in
the  comoving volume  contained inside the horizon:
\begin{equation}
N_{\rm s}
  =  \int dL ~\int_0^{z_h(L)} dz~v(z)~ n(L,z )
\label{eq:ndet}
\end{equation}
A  model  for $n(L,z)$ to describe the  luminosity distribution
and cosmic  evolution of the sources allows
to  predict the fraction
of the extra--galactic associated   with the resolved  flux, 
and the   corresponding  number of  sources.
Here there is no  space for a full discussion, 
but it  may be instructive to consider a simple toy model
where all sources   have identical   luminosity 
(that is $n(L,z) = n \; \delta [L - \overline{L}(z)]$).
In this  model, is the (unique) luminosity 
$L$ of the sources  at the present  epoch 
is not too large,
(that is for  $L \aprle  L^*$  with $L^*$  given 
in (\ref{eq:lstar})), for 
a {\em  fixed} total flux  for the  ensemble of 
sources,  the  number of  objects that can be
resolved is (for $\alpha\simeq 2$):
\begin{eqnarray}
N_{\rm s}
 & ~\simeq  ~ &   
\frac{\sqrt{4 \pi}}{3} \,
\frac{1}{\xi~ \log(10)}
\; \frac{H_0}{c} 
\frac{K_\nu \, \sqrt{L_{\rm decade}}}
  {(\Phi_{\rm min} ~E_{\rm min})^{3/2}}
\nonumber 
\\
&  ~\simeq ~ &  10 
~\left ( K_{-9} \right)
 \, \left ( \Phi^{\rm min}_{-11} \right)^{-\frac{3}{2}} 
 \, \left ( L_{45} \right)^{\frac{1}{2}} 
 \, \frac{2.5}{\xi}
\end{eqnarray}
where  $K_\nu$  is  the coefficient of the  diffuse 
$\nu$ flux  ($K_{-9}$ is  in units  $10^{-9}$~GeV/(cm$^2$~s~sr)),
$L_{\rm dec}$ is the power of an individual  source per 
energy decade ($L_{45}$ is in units of $10^{45}$~erg/s),
and $\Phi_{\rm min}$ is the minum  flux  above  energy 
$E_{\rm min} = 1$~TeV  for  source identification
($\Phi_{-11}$ is in units $10^{-11}$~(cm$^{2}$s)$^{-1}$).
It is  easy to understand the scaling   laws.
When  the luminosity  of the source  increases
the radius of  the  source horizon   
(for $L$ not too large)  grows  $\propto \sqrt{L/\Phi_{\rm min}}$
and the corresponding  volume   grows as
 $V \propto (L/\Phi_{\rm min})^{3/2}$, while  (for a fixed
total  flux) the number  density  of the sources 
is $n\propto K_\nu \, L^{-1}$,
 therefore the number  of  detectable sources
is: $N_{\rm s} \propto n \, V 
\propto K_\nu \;L^{1/2} \; \Phi_{\rm min}^{-3/2}$.

Similarly, the  ratio  of the resolved  to the
total  $\nu$  flux  can be  estimated as:
\begin{eqnarray}
\frac{\phi_{\rm resolved}} {\phi_{\rm total}} 
 & \simeq  &  
\frac{1}{4 \pi} 
~\frac{1}{\xi~\log(10)} 
~\frac{H_0^2}{c^2}  
~ \frac{L_{\rm dec}}{\Phi_{\rm min} ~E_{\rm min}} 
\\
&  \simeq &  0.005~
~\left ( \Phi^{\rm min}_{-11} \right)^{-1} 
~\left ( L_{45} \right)
 \, \frac{2.5}{\xi}
\end{eqnarray}
The scaling $\propto L/\Phi_{\rm min}$ also  easily follows
from the assumption  of an euclidean  near universe.

The bottom line of this discussion, is  that 
it is very likely  that 
the ensemble of all extra--galactic sources  will give its largest
contribution  as an unresolved , isotropic contribution, with only
 a small fraction  of this total  flux resolved in the contribution
of few  individual point  sources.
The  number of   the detectable extra--galactic point
sources  will   obviously   grow   linearly 
with  total extra--galactic flux,  but  also depends crucially 
on   the luminosity  function and  cosmological evolution
of the $\nu$ sources.  If a reasonable fraction
of the   individual  sources are
sufficiently powerful ($L \aprge 10^{45}$~erg/s),   
an interesting number of objects can be  detected as  point sources.
Emission for  blazars is a  speculative but very exciting possibility
(see fig.~\ref{fig:horizon}).

\section{GZK neutrinos}
For  lack of space we cannot discuss here 
the  neutrinos  of ``GZK'' origin 
and other  speculative  sources \cite{Semikoz:2003wv}.
This field is  of great interest, 
and  is  in many ways  complementary
to the science that  can be performed  with 
km$^3$  telescopes.
GZK  neutrinos  
(see fig.\ref{fig:cr_nu}) 
have  energy  typically 
in the 10$^{18}$--10$^{20}$~eV, and the 
predicted fluxes are  so  small that
km$^3$ detectors can only be 
very marginal,  and larger  detector masses 
and new detection methods  are in order.
Several interesting  ideas are being developed
(for a review see \cite{Learned:2003}),
these include acoustic  \cite{Saltzberg}, 
radio  \cite{Falcke} and Air Shower 
\cite{Letessier-Selvon,euso-nu} detection.

The GZK neutrinos are a guaranteed source, and 
their measurement  carry  important information
on the maximum energy and cosmic history
of the ultra high energy cosmic rays.
Perhaps the  most promising detection technique
uses radio  detectors on balloons. 
An  18.4 days  test flight of the ANITA experiment \cite{anita-lite}
(see fig.~\ref{fig:cr_nu}) has  obtained  the best limits  on
the diffuse flux of neutrinos at very high energy.

As photon  astronomy is  articulated  in different
fields according to the range of  wavelength observed,
for neutrino  astronomy one can  already
see the formation  of (at least)  two  different  subfields:
the ``km$^3$ neutrino  Science'' that aims at the  study of
$\nu$  in the 10$^{12}$--10$^{16}$~eV   energy range,
and ``Ultra High Energy neutrino Science''  that 
studies $\nu$  above  10$^{18}$~eV,
with the detection of  GZK neutrinos as the primary goal.

\section{Conclusions and Outlook}
The potential of the km$^3$ telescopes to open 
the extraordinarily interesting new  window
of high energy  neutrino astronomy  is good.
The closest thing to a  guaranteed  $\nu$ source
are the young SNR  observed  by HESS  in TeV photons.
The  expected  $\nu$ fluxes from these sources are
probably above  the  sensitivity of a  km$^3$  telescopes
in the northern hemisphere.
Other promising   galactic $\nu$ sources  are the Galactic Center
and $\mu$Quasars.  Blazars  and GRB's are also intensely discussed
as extragalactic $\nu$ sources.
The combined $\nu$ emission   of all extragalactic  sources should also
be detectable  as a diffuse flux distinguishable from
the atmospheric $\nu$ foreground.
Because of the important  astrophysical uncertainties,
the clear observations of astrophysical neutrinos
in the km$^3$ telescopes is  however not fully guaranteed.

There are currently  plans to build two different  instruments
of comparable performances (based on the water Cherenkov technique
in water and ice) in  the  northern and  southern hemisphere.
Such instruments  allow  a complete coverage of the celestial sphere.
This  is a very important scientific  goal,    if the detector
sensitivity is sufficient to perform  interesting observations.
If the deployment  of one detector  anticipates the
second one, it is necessary  to be prepared to modify
the  design of the second one, on the basis  of the lessons
received. This  is particularly important 
if the first  observations  give no (or  marginal) evidence
for  astrophysical neutrinos, indicating 
the need of an enlarged acceptance.

A  personal ``guess''  about the  most
likely outcome  for  the operation of 
the km$^3$  telescopes, is that they will
play  for  neutrino  astronomy
a role similar to  what the first   $X$--ray  rocket  of
Rossi and Giacconi\cite{giacconi,giacconi-first} played
for $X$--ray  astronomy in 1962.
That  first   glimpse of  the
X--ray sky revealed one  single point source, 
the AGN Sco-X1 (that a the moment was in a
high state of  activity),
and  obtained  evidence for an  isotropic
$X$--ray light glow of the  sky.
Detectors of higher 
sensitivity   (a factor 10$^4$ improvement in 40 years)
soon started to  observe a  large   number
of  sources belonging to  different classes.

Even in the most optimistic  scenario, the 
planned km$^3$ telescopes will
just ``scratch  the  surface'' of  the   rich  science
that the neutrino messenger will carry.
To explore  this field  it will be obviously necessary
to  develop higher acceptance detectors, and it is not
too soon to think in this  direction.

\vspace{0.3cm}
\noindent {\bf Acknowledgments}. 
I'm very grateful to Emilio Migneco, Piera Sapienza, Paolo Piattelli
and the colleagues from Catania for the organization  of a very fruitful
workshop.  I have  benefited  from  conversations
with  many colleagues. Special thanks to Tonino Capone
and  Felix Aharonian.

\end{document}